\documentclass[prb,aps,twocolumn,epsf,showpacs]{revtex4}
                        
\include{epsf}

\begin{document}                                    
\title{In Search of the Vortex Loop Blowout Transition \\for a                  
type-II Superconductor in a Finite Magnetic Field}                 
                                                                
\author{P. Olsson}                                                             
\affiliation{Department of Theoretical Physics, Ume{\aa} University, 901 87
Ume{\aa} Sweden}                                    
                                                                                
\author{S. Teitel}                                         
\affiliation{Department of Physics and Astronomy, University    
of Rochester, Rochester, NY 14627}                         
                                                                 
\date{\today}                                                             
                                                                            
\begin{abstract}                                                                

The 3D uniformly frustrated XY model is simulated to search for
a predicted ``vortex loop blowout'' transition within the vortex
line liquid phase of a strongly type-II superconductor in an applied
magnetic field.  Results are shown to strongly depend on the precise
scheme used to trace out vortex line paths.  While we find evidence for a 
transverse vortex path percolation transition, no  signal of this 
transition is  found in the specific heat.

\end{abstract}

                                                         
\maketitle                                          

\section{Introduction}

In pure extreme type-II superconductors, such as the high $T_c$
superconductors, the Abrikosov vortex line lattice melts via a sharp 
first order phase transition\cite{R1} into a vortex line liquid as the temperature 
is increased above a critical $T_m$.  The properties of this vortex line liquid 
phase have been the subject of considerable investigation.
Theoretical arguments\cite{R2} and early simulations\cite{R3,R3.1,R3.2} suggested that the vortex
line liquid might retain superconducting phase coherence parallel to 
the applied magnetic field, within some temperature interval above 
$T_m$.  Later, better converged simulations\cite{R4} found that phase 
coherence is simultaneously lost in all directions upon melting.

Subsequently, Te\v{s}anovi\'{c}\cite{R5} has proposed that, for small magnetic
fields, there still remains
a sharp thermodynamic phase transition at a temperature $T_\Phi$
within the vortex liquid state, associated with diverging fluctuations 
of closed vortex loops, such as drive the superconducting transition 
in the zero magnetic field case.  Considering the limit of infinite 
penetration length $\lambda$, Te\v{s}anovi\'{c} proposed that,  in a 
finite field, the 
fluctuations of the magnetic field induced vortex lines act to screen
the interactions of thermally excited closed vortex loops, in the 
same way that
magnetic field fluctuations screen the vortex loop interactions of
a finite $\lambda$ model in zero applied magnetic field.  Pursuing
this argument, he predicted that the proposed vortex loop blowout
transition at $T_\Phi$
may be an inverted XY transition, as is the case of the zero field
Meissner transition for the finite $\lambda$ model. 

Following Te\v{s}anovi\'{c}'s predictions, Sudb{\o} and 
co-workers\cite{R6,R7,R8} have
carried out numerical simulations of the three dimensional (3D) 
uniformly frustrated XY model
of a type-II superconductor.  They claim to find evidence for
Te\v{s}anovi\'{c}'s transition, which they associate with the formation
of a vortex line path that percolates entirely around the system
in the direction transverse to the magnetic field.  

Most recently, measurements\cite{R9} on high purity YBa$_2$Cu$_3$O$_7$ 
(YBCO) single crystals have found a step 
like anomaly in the specific heat at a temperature highter than 
the melting $T_m$, reminiscent
of an inverted mean field transition.  It has been argued\cite{R9} that this
feature may be evidence for Te\v{s}anovi\'{c}'s transition $T_\Phi$.

In order to further investigate this issue, we have carried out new 
simulations on the 3D uniformly frustrated XY model, both repeating 
the approach of Sudb{\o} and co-workers, and measuring new quantities
that make a more direct test of Te\v{s}anovi\'{c}'s theory.  After 
correcting certain inconsistencies in the earlier numerical work, we 
show that whether or not one finds indications of a vortex loop 
blowout transition depends crucially on how one chooses to resolve 
vortex line paths at points where two or more lines intersect.  Making
the choice that favors the blowout interpretation, we find the critical 
exponent $\nu\simeq 1$, rather than the value $\sim 2/3$ expected for 
an inverted 3D XY transition.
Finally, we make high precision measurements of the specific 
heat, in search of a thermodynamic signature for a blowout 
transition, but no such signature is found.

\section{Model}

The model that we use is the 3D uniformly frustrated XY 
model\cite{R3,R3.1,R3.2,R4,R6,R7,R8,R10} which models a type-II superconductor 
in the limit of infinite magnetic
penetration length, $\lambda\to\infty$, and is given by the Hamiltonian, 
\begin{equation}
    {\cal H}[\theta_{i}] = -\sum_{i, \mu}J_{\mu}\cos 
    (\theta_{i+\hat\mu}-\theta_{i}-A_{i\mu})\enspace.
\label{eqH}
\end{equation}
Here $i$ are the nodes of a cubic grid of sites, $\mu =  
x, y,$ $z$, are the directions of the grid axes,
and the sum is over  all nearest neighbor bonds of the grid.
$\theta_{i}$ is the phase angle of the superconducting 
wavefunction on site $i$, 
$A_{i\mu}=(2\pi/\Phi_{0})\int_{i}^{i+\hat\mu}{\bf A}\cdot d{\bf r}$
is given by the integral of the magnetic vector potential ${\bf A}$ across
the bond at site $i$ in direction $\hat\mu$, and $\Phi_{0}=hc/2e$ is 
the flux quantum.  The argument of the cosine is the gauge
invariant phase angle difference across the bond.
The circulation of the $A_{i\mu}$ around any 
plaquette of the grid is equal to $2\pi$ times number of flux quanta 
of magnetic field penetrating the plaquette.  We take the magnetic 
field, ${\bf B}=\nabla\times{\bf A}$, uniform and parallel to the 
$\hat z$ axis, with a fixed density of flux quanta $f=Ba^{2}/\Phi_{0}$
per plaquette of area $a^{2}$.  The couplings $J_{\mu}$ we take to 
model an anisotropic system, with $J_{x}=J_{y}\equiv J_{\perp}$, and 
$J_{z}\le J_{\perp}$.

Simulations are carried out varying the $\theta_i$ according to a 
usual Monte Carlo scheme; the $A_{i\mu}$ are held fixed.  That the
$A_{i\mu}$ do not fluctuate, and that they give a uniform magnetic field, 
are the consequences of the $\lambda\to\infty$ approximation.  Simulations
are carried out on $L_z\times L_\perp^2$ cubic grids, using periodic
boundary conditions.   Except where otherwise noted, 
our runs are typically for $1/4$ to $1/2\times 10^6$ Monte Carlo 
passes through the entire lattice.

While we 
simulate in the phase angle degrees of freedom, $\theta_i$, our 
interest will be in the behavior of the vorticity in these phase 
angles.  Let $s$ denote the {\it dual} sites of the original grid; 
these are the sites at the centers of the unit cells of the grid.
Denote by  $(s,\mu)$ the plaquette which is the face of the unit cell 
centered on 
dual site $s$ with normal in the $\hat\mu$ direction, $\mu = x,y$, $z$.
We define the integer vorticity $n_{s\mu}$ piercing 
plaquette $(s,\mu)$ by computing the circulation of the
gauge invariant phase angle differences around the plaquette,
\begin{equation}
    \sum_{(s,\mu)} [\theta_{i+\hat\sigma}-\theta_i - A_{i\sigma}] 
    = 2\pi (n_{s\mu} -f\delta_{z\mu})\enspace,
\end{equation}
where the sum is counterclockwise around all bonds forming the boundary 
of plaquette $(s,\mu)$, and the gauge invariant phase angle 
differences are restricted to the interval $[-\pi,\pi)$.  
In a constant magnetic field, the condition that the total energy
density remains finite can be shown to yield the ``neutrality'' 
constraint (see section III.A),
\begin{equation}
    \sum_{s}n_{s\mu}=fL_{\perp}^{2}L_{z}\delta_{z\mu}\enspace,
    \label{eqNeutral}
\end{equation}
i.e. the total vorticity piercing any plane at constant $z$ is 
$fL_{\perp}^{2}$; these are the magnetic field induced vortex lines.
The total vorticity in the transverse directions $x$ and $y$ is zero.

Taking the vorticity $n_{s\mu}$ as the directed
bond of the dual grid, emanating from site $s$ in direction $\hat\mu$,
the vorticity so defined is divergenceless, 
forming continuous lines that, due to the periodic boundary
conditions, must ultimately close upon themselves.  We will label 
such a closed vortex path by the index $\alpha$, and
define the vector ${\bf R}_\alpha$ as the net displacement one travels 
upon following the path $\alpha$ from a given starting point until 
returning back to that point as the line closes back on itself.
If ${\bf R}_\alpha=0$, then the vortex line path is a closed loop 
of finite extent 
that exists as a thermal fluctuation.  If $R_{z\alpha}=mL_{z}$, $m$ 
integer, then such a
vortex line path represents $m$ of the $fL_{\perp}^{2}$ field induced 
vortex lines; these $m$ lines are mutually connected to each
other via the periodic boundary conditions in the $z$ 
direction.\cite{R3.1} 
For $m>1$, we can say that the $m$ field induced lines are
geometrically entangled with each other.
If $R_{x\alpha}=mL_{\perp}$ or $R_{y\alpha}=mL_{\perp}$, then the 
vortex line path winds $m$ times around the system {\it transversely} to 
the applied magnetic field.  We will be particularly interested in 
vortex line paths for which $R_{z\alpha}=0$, but $R_{x\alpha}$ or $R_{y\alpha}\ne 0$.
Henceforth, we will refer to as the ``lines'', 
the set of vortex line paths $\{\alpha\}$ for which all $R_{z\alpha}>0$; 
these are the field induced vortex lines.  All 
other vortex paths we will refer to as the ``loops''.

In order to trace out vortex line paths, one needs to know how to 
treat intersections.  An intersection is when there is more than one vortex 
line entering and exiting a give unit cell of the grid, and it is 
therefore ambiguous which entering segment to connect to which 
exiting segment.  It was previously shown\cite{R11} by one of us that the method 
chosen to resolve such intersections can have a dramatic effect on the 
statistics of closed thermally excited loops in the zero field $f=0$ 
model.  Here, for the $f>0$ model, we consider two different schemes,
which we henceforth refer to as method (i) and method (ii):

(i) At each intersection we choose randomly, with equal probability, 
which entering segment connects to which exiting segment.  In the 
$f=0$ model this scheme was found to give results closest to 
theoretical expectations.\cite{R11}

(ii) Motivated by Sudb{\o} and co-workers,\cite{R6,R7,R8} we 
first search\cite{R11.1} through all possible connections to find a path 
$\alpha$ with $R_{z\alpha}=0$ and $R_{x\alpha}$ or 
$R_{y\alpha}\ne 0$.  
Such a path winds around the system transverse to the field, without
ever winding around the system parallel to the field.
If one such path is found, it is selected as a path $\alpha^\prime$
contributing to the ``loops'', and we then repeat the proceedure
applied to all remaining vortex paths.  When all such transverse paths 
are found, the remaining vortex line intersections are 
resolved randomly, as in method (i).

Using either method (i) or method (ii) we thus decompose the vorticity
of any given configuration into disjoint closed 
vortex line paths, consisting of a set $\{\alpha\}$ of ``lines'' and 
a set $\{\alpha^{\prime}\}$ of ``loops''.

\section{Winding of Field Induced Vortex Lines}

We first attempt a direct test of Te\v{s}anovi\'{c}'s theory
of the $T_\Phi$ transition within the liquid phase.  A 
summary of his arguments for the existence of this transition 
is as follows.

\subsection{Summary of Te\v{s}anovi\'{c}'s Theory}

First, a duality transformation\cite{R12,R12.1,R12.2} from the XY 
model of Eq.\,(\ref{eqH}) gives the interaction between 
vortices as,
\begin{equation}
    {\cal H}[n_{s\mu}]={1\over 2}\sum_{s,s^{\prime},\mu} 
    [n_{s\mu}-f\delta_{z\mu}]V^{\mu}({\bf r}_{s}-{\bf 
    r}_{s^{\prime}})[n_{s^{\prime}\mu}-f\delta_{z\mu}]
\label{eqV}
\end{equation}
where $V^{\mu}({\bf r})$ is the appropriate anisotropic 
generalization of the Coulomb interaction, with Fourier transform 
$V_{q}^{\mu}\sim q^{-2}$.  It is this singularity of $V_{q}^{\mu}$
as $q\to 0$ that yields the constraint of Eq.\,(\ref{eqNeutral}).

Next, one imagines decomposing the total 
vorticity of the system into lines and loops,
\begin{equation}
    n_{s\mu}= n_{s\mu}^\mathrm{lines}+ n_{s\mu}^\mathrm{loops}\enspace.
\end{equation}
If we define,
\begin{equation}
    b_{s\mu}\equiv n_{s\mu}^\mathrm{lines}-f\delta_{z\mu}
\label{eqb}
\end{equation}
then $\sum_{s}\langle b_{s\mu}\rangle=0$ and the Hamiltonian of Eq.\,(\ref{eqV}) can
be rewritten as,
\begin{equation}
    {\cal H}={1\over 2}\sum_{s,s^{\prime},\mu} 
    [n_{s\mu}^\mathrm{loops}-b_{s\mu}]V^{\mu}({\bf r}_{s}-{\bf     
    r}_{s^{\prime}})[n_{s^{\prime}\mu}^\mathrm{loops}-b_{s^{\prime}\mu}]\enspace.
\label{eqLoop}
\end{equation}

Te\v{s}anovi\'{c} then argues that a coarse graining of vortex 
fluctuations, in the vortex line liquid phase, leads to an effective 
hydrodynamic Hamiltonian on long length scales 
which has the same interaction piece 
as Eq.\,(\ref{eqLoop}), but which has a new additive term proportional to
$\sum_{s\mu}b_{s\mu}^{2}$.  The resulting long length scale Hamiltonian 
then has exactly the same form as that of a zero field superconductor 
with thermally fluctuating vortex loops, $n_{s\mu}^\mathrm{loops}$, {\it and} 
a thermally fluctuating magnetic field $b_{s\mu}$ whose average is zero, i.e. 
the zero field superconductor with a finite penetration length 
$\lambda$.  In other words, in this {\it infinite} $\lambda$ theory at {\it 
finite} magnetic field, the long wave length 
fluctuations of the field induced vortex lines $n_{s\mu}^\mathrm{lines}$ 
screen the interaction between the vortex loops $n_{s\mu}^\mathrm{loops}$ in 
exactly the same manner as magnetic field fluctuations screen the 
interactions between vortex loops in a {\it finite} $\lambda$ model 
at {\it zero} magnetic field.

The Meissner transition at $T_{c}$ 
in the zero field, finite $\lambda$, model is an 
{\it inverted} 3D XY transition.\cite{R12}  The high temperature phase 
$T>T_{c}$ has 
vortex loops on all length scales and breaks a global $U(1)$ symmetry 
associated with a {\it disorder} parameter;\cite{R12.3} the low temperature phase 
$T<T_{c}$ has no vortex loops on sufficiently long length scales.  The 
correlation length $\xi$ and renormalized 
magnetic penetration length $\lambda_{R}$ both diverge\cite{R13} as $\sim 
|t|^{-\nu}$, with $\nu\simeq 2/3$ and $t\equiv T-T_{c}$.

We have earlier carried out numerical simulations\cite{R13} of 
this zero field, finite $\lambda$, Meissner transition.
We demonstrated that, in this model,
magnetic field fluctuations obey the finite size scaling relation,
\begin{eqnarray}
    F(t,q,L)&\equiv&\langle b_\mu(q\hat\nu)b_\mu(-q\hat\nu)\rangle/L^3 
    \nonumber \\    &\sim&    L^{-1}F(tL^{1/\nu}, qL, 1)
\label{eqScal}
\end{eqnarray}
where in the above $\hat\mu\perp\hat\nu$ and
$b_\mu(q\hat\nu)=\sum_s e^{-iq\hat\nu\cdot{\bf r}_s}b_{s\mu}$ is the Fourier transform 
of of the magnetic flux density $b_{s\mu}$.  As $L\to\infty$, and
$q\to 0$,
\begin{equation}
    F(t,0,\infty) \sim \left\{ 
    \begin{array}{ll}
	0 & t<0\\
	1/\xi & t>0
    \end{array}
    \right.\enspace,
\label{eqScal1}
\end{equation}
hence $F(t,0,\infty)$ vanishes below the transition, and increases 
continuously from zero as one goes above the transition.    
    
In the present case of a finite magnetic field, if 
Te\v{s}anovi\'{c}'s mapping is correct, the Meissner transition
$T_{c}$ becomes the transition $T_{\Phi}$ within the vortex line
liquid phase, and we expect the exact same scaling 
as in Eq.\,(\ref{eqScal}) above, 
when applied to the quantity $b_{s\mu}$ defined in 
Eq.\,(\ref{eqb}).  Taking the limit of $q\to 0$ in Eq.\,(\ref{eqScal}), 
and applying to systems with fixed aspect ratio $L_z=g 
L_\perp$, we expect the scaling,
\begin{equation}
    \big\langle (\sum_s b_{s\mu})^2\big\rangle/L_\perp^3 \sim 
    L_\perp^{-1} f(tL_{\perp}^{1/\nu})\enspace,
\label{eqScal2}
\end{equation}
where $t\equiv T-T_{\Phi}$ and $f(x)=F(x,0,1)$.

For the directions $\hat\mu = \hat x$ or $\hat y$, 
\begin{equation}
    \sum_{s}b_{s\mu}=\sum_{s}n_{s\mu}^\mathrm{lines}\equiv W_{\mu}L_{\perp}
\label{eqWp}
\end{equation}
is the net vorticity of the magnetic field induced vortex lines
in the transverse direction $\hat\mu$.
The two dimensional vector ${\bf W} = (W_{x},W_{y})$ defined above is the 
integer valued ``winding number''
that counts the net number of times the field induced vortex lines
wind around the system in the transverse directions $\hat x$ and 
$\hat y$.  
If $\{\alpha\}$ is the set of vortex line paths that define the 
field induced vortex lines $n_{s\mu}^\mathrm{lines}$, and
${\bf R}_{\alpha}$ is the net displacement along path $\alpha$ as defined 
earlier, then $\sum_{\alpha}{\bf R}_{\alpha\perp}={\bf W}L_{\perp}$,
where ${\bf R}_{\perp}\equiv (R_{x},R_{y})$.  We thus
expect from Eq.\,(\ref{eqScal2}) the finite size scaling,
\begin{equation}
    \langle W^{2}\rangle \sim f(tL_{\perp}^{1/\nu})\enspace.
\label{eqScal3}
\end{equation}

Note, the neutrality condition of Eq.\,(\ref{eqNeutral}) implies that 
the {\it total} transverse vorticity in the system must always 
vanish.  For ${\bf W}\ne 0$, it is therefore necessary that any such 
winding of the field induced lines is exactly canceled out by an 
equal and opposite transverse winding of the loops.  In the 
thermodynamic limit, $L_{\perp}\to\infty$, Eq.\,(\ref{eqScal1})
implies that $\langle W^{2}\rangle=0$ for $T<T_{\Phi}$, and $\langle 
W^{2}\rangle$ increases continuously from zero as one increases $T>T_{\Phi}$.
The proposed transition at $T_{\Phi}$ is thus associated with the appearance 
of infinite transverse loops (see following Section IV).

Another interpretation of the $T_{\Phi}$ transition  follows
from the ``two dimensional (2D) boson'' model\cite{R14} of interacting vortex lines, 
in which the field induced vortex lines are viewed as the world lines 
of two dimensional bosons traveling down the imaginary time axis.
For $T<T_{\Phi}$ where $\langle W^2\rangle=0$, 
the field induced vortex lines 
behave like {\it charged} two dimensional bosons,\cite{R2,R12.2} with a long range 
retarded Coulomb interaction.  In the vortex line liquid, $T_{m}<T<T_{\Phi}$, 
where phase coherence is lost parallel to the applied magnetic field, 
the analog 2D bosons are in a {\it charged} superfluid state.
For $T>T_{\Phi}$, where $\langle W^2\rangle\ne0$, screening by the 
infinitely large loops $n_{s\mu}^\mathrm{loops}$ results in an effective 
short ranged interaction between the field induced lines.  In this 
case the winding number squared $\langle W^{2}\rangle$ is proportional to 
the superfluid density of what is now an {\it  uncharged} superfluid.
Thus $T_{\Phi}$ corresponds to a transition between a charged 
superfluid and an uncharged superfluid in the analog 2D boson theory.
Equivalently, if one considers the quanta that mediate the 
interaction between the analog 2D bosons, the transition is from 
massless quanta for $T<T_{\Phi}$ to massive quanta for $T>T_{\Phi}$.

To arrive at Eq.\,(\ref{eqScal3}), we considered the transverse 
components of Eq.\,(\ref{eqScal2}).  However, we can also consider the 
parallel, $\hat\mu=\hat z$ component.  Now,
\begin{equation}
    \sum_{s}b_{sz}=\sum_{s}n_{sz}^\mathrm{lines}-fL_{\perp}^{2}L_{z}\equiv 
    W_{z}L_{z}\enspace,
\end{equation}
and so we expect the scaling
\begin{equation}
    \langle W_{z}^{2}\rangle \sim f_{z}(tL_{\perp}^{1/\nu})\enspace.
    \label{eqscalz}
\end{equation}
If $\{\alpha\}$ is the set of vortex line paths that define the 
field induced vortex lines $n_{s\mu}^\mathrm{lines}$, and
${\bf R}_{\alpha}$ is the net displacement along path $\alpha$ as defined 
earlier, then $\sum_{\alpha}R_{\alpha z}=(fL_{\perp}^{2}+W_{z})L_{z}$.
Thus $W_{z}$ gives the number of ``lines'' in excess of the
average value $fL_{\perp}^{2}$ set by the applied magnetic field.  The 
neutrality condition of Eq.\,(\ref{eqNeutral}) requires that when 
$W_{z}>0$, there must be an equal and opposite parallel winding of the 
loops $n_{s\mu}^\mathrm{loops}$.  As $L\to\infty$, we have
$W_{z}=0$ for $T<T_{\Phi}$, and $W_{z}>0$ 
for $T>T_{\Phi}$.  Thus a transition at $T_{\Phi}$ should be 
characterized by fluctuations in the number of field induced lines
and by the appearance of infinite parallel loops directed
opposite to the direction of the applied magnetic field.

\subsection{Numerical Results}

To test the above predictions, we have simulated the 3D 
uniformly frustrated XY model of Eq.\,(\ref{eqH}) using 
a vortex density $f=1/20$, anisotropy $J_{z}/J_{\perp}=0.02$, 
and aspect ratio $L_{z}/L_{\perp}=1$, for $L_{\perp}=10,$ $20$,
$30$, $40$ and $60$.  
For these parameters, the vortex lattice melting temperature
is $T_{m}\simeq 0.24 J_{\perp}$, and the zero field critical
temperature is $T_{c0}\simeq 1.14 J_{\perp}$. 
We compute the transverse winding ${\bf W}$ of 
the field induced lines, defined by Eq.\,(\ref{eqWp}),
using both method (i) and method (ii) to 
decompose each configuration into ``lines'' and ``loops''.
According to the scaling equation (\ref{eqScal3}), we expect
that plots of $\langle W^{2}\rangle$ vs. $T$ for different sizes
$L_{\perp}$ should all intersect at the common point
$t=0$, or $T=T_{\Phi}$.  

\begin{figure}
\epsfxsize=3.2truein
\epsfbox{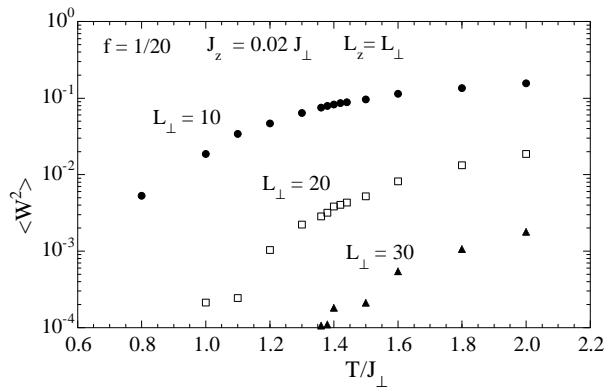}
\caption{Semilog plot of winding $\langle W^{2}\rangle$ vs. $T/J_{\perp}$ for
$L_{\perp}=10$, $20$, and $30$, with vortex density $f=1/20$,
anisotropy $J_z=0.02 J_\perp$, and aspect ratio $L_z=L_\perp$.
$\langle W^{2}\rangle$ is computed using method (i), i.e. random 
connections at intersections.  $\langle W^{2}\rangle$ steadily 
decreases as $L_{\perp}$ increases, over the entire temperature range.
\label{f1}}
\end{figure}

In Fig.\,\ref{f1} we show a semilog plot of $\langle W^{2}\rangle$
vs. $T/J_{\perp}$ using method (i) (random reconnections at intersections)
for $L_{\perp}=10$, $20$, and $30$.  We see that there is clearly
no common intersection point of the curves.  As $L_{\perp}$ 
increases, $\langle W^{2}\rangle$ decreases uniformly over the entire 
temperature range.  This is in qualitative agreement with earlier 
computations 
of $\langle W^{2}\rangle$ by one of us (see Fig.~15 of 
Ref.\onlinecite{R3.1}). 
For $L_{\perp}=60$, we have 
found no net transverse winding of the field induced lines at all, i.e.
for the length of our simulation we had
${\bf W}=0$, for the temperature range $1.36\le T/J_{\perp}\le 1.44$.

\begin{figure}
\epsfxsize=3.2truein
\epsfbox{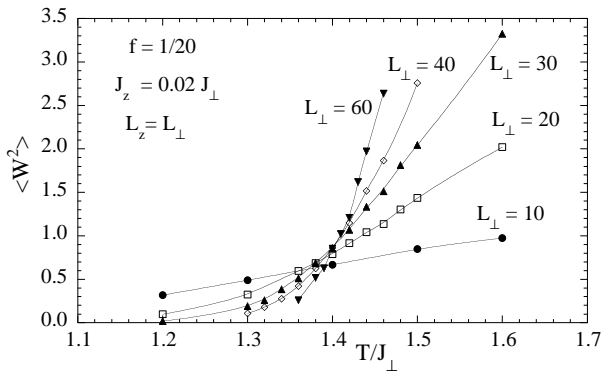}
\caption{Plot of winding $\langle W^{2}\rangle$ vs. $T/J_\perp$ for
$L_{\perp}=10$, $20$, $30$, $40$, and $60$, with vortex density $f=1/20$,
anisotropy $J_z=0.02 J_\perp$, and aspect ratio $L_z=L_\perp$.
$\langle W^{2}\rangle$ is computed using method (ii), i.e. first find 
all percolating transverse loops.  Curves of $\langle W^{2}\rangle$
intersect at a common point, locating $T_{\Phi}\simeq 1.4 J_{\perp}$.  
Solid lines are guides to the eye only.
\label{f2}}
\end{figure}

Next, in Fig.\,\ref{f2}, we show the same quantities but now using 
method (ii) (search first for maximal transverse loops), for 
$L_{\perp}=10,$ $20$, $30$, $40$ and $60$.  We see 
that as $L_{\perp}$ increases, the curves do seem to approach a common 
intersection point, giving a $T_{\Phi}\simeq 1.4 J_{\perp}$. 
Note that this $T_{\Phi}$
is {\it above} the zero field critical temperature 
$T_{c0}\simeq 1.14 J_{\perp}$.

\begin{figure}
\epsfxsize=3.2truein
\epsfbox{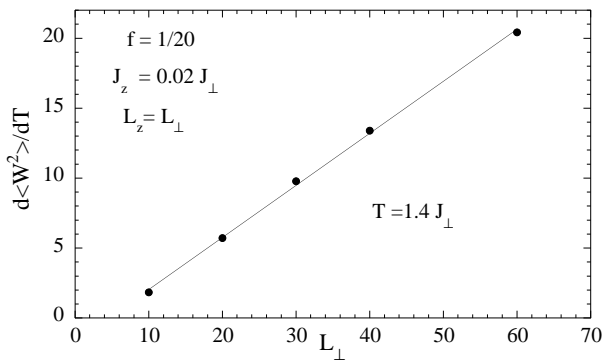}
\caption{Plot of winding slopes $d\langle W^{2}\rangle/dT$ vs. $L_\perp$ 
at the estimated crossing temperature of Fig.\,\ref{f2}, 
$T=1.4J_\perp$, for
$L_{\perp}=10$, $20$, $30$, $40$, and $60$, with vortex density $f=1/20$,
anisotropy $J_z=0.02 J_\perp$, and aspect ratio $L_z=L_\perp$.
The solid line is the best linear fit to the data.  The good fit 
suggests the critical exponent $\nu\simeq 1$.
\label{f2.1}}
\end{figure}

From Eq.\,(\ref{eqScal3}), we expect that the slopes of 
these curves at $T_{\Phi}$ should scale with system size as, $d\langle 
W^2\rangle/dT\sim L_\perp^{1/\nu}$.  Fitting each of the curves of $\langle 
W^2\rangle$ to a cubic polynomial in $T$, we compute their derivatives at the
intersection point $T=1.4J_\perp$, and plot the results vs. $L_\perp$ 
in Fig.\,\ref{f2.1}.  We see that the slopes, to an excellent 
approximation, scale linearly with $L_\perp$, thus suggesting a
critical exponent $\nu\simeq 1$.  On closer inspection, the data in 
Fig.\,\ref{f2.1} show a small systematic downwards curvature about the 
linear fit; however this curvature can be removed by assuming a 
slightly higher critical temperature of $T=1.403 J_\perp$.  Note that
this value of $\nu\simeq 1$ is {\it larger} than the predicted value of $2/3$.

\begin{figure}
\epsfxsize=3.2truein
\epsfbox{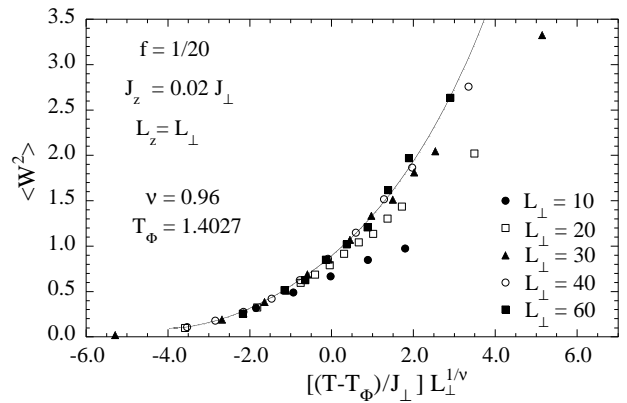}
\caption{Scaling collapse of data of Fig.\,\ref{f2}. $\langle W^{2}\rangle$ 
plotted vs. $[(T-T_{\Phi})/J_{\perp}]L_{\perp}^{1/\nu}$, for
$L_{\perp}=10$, $20$, $30$, $40$, and $60$, with vortex density $f=1/20$,
anisotropy $J_z=0.02 J_\perp$, and aspect ratio $L_z=L_\perp$.
Data is fit to a polynomial expansion of Eq.\,(\ref{eqScal3}), and
$T_{\Phi}\simeq 1.4027 J_{\perp}$ and $\nu\simeq 0.96$ determined from the fit.
Only data from $L_{\perp}=40$ and $60$ are used in the fit, although 
data from all sizes are shown in the plot. 
The solid line is the fitted polynomial curve.
\label{f3}}
\end{figure}

As an alternative method to compute the critical behavior, we can
take the scaling equation (\ref{eqScal3}), expand the scaling
function $f(x)$ as a polynomial for small $x$, and do a nonlinear
fitting to  the data to determine the unknown polynomial coefficients, 
$T_{\Phi}$, and $\nu$.  To obtain the best fit we use a 4th order
polynomial and fit only the data
from the two largest sizes, $L_{\perp}=40$ and $60$.
The results give $T_{\Phi}\simeq 1.403 J_{\perp}$ and $\nu\simeq 0.96$,
in agreement with the earlier estimates.  In Fig.\,\ref{f3}
we show the scaling collapse that results from this polynomial fit.
There are systematic deviations from the fitted curve on the $T>T_{\Phi}$ 
side, though these appear to decrease as $L_{\perp}$ increases.  

Next we consider the excess parallel winding of the field induced 
lines $W_z$.  As discussed earlier, Te\v{s}anovi\'{c}'s theory 
predicts a scaling of $\langle W_z^2\rangle$ as in  
Eq.\,(\ref{eqscalz}).  To determine $W_z$ we count the winding of 
vortex line paths $\{\alpha\}$ that wind negatively in the $z$ direction, i.e. have 
a net displacement of $R_{\alpha z}=-m_\alpha L_z$, with $m_\alpha$ a 
positive integer (${\bf R}_\perp$ may have any value for such paths).
Since such negative parallel windings must be compensated for by 
excess field ``lines'', we have $W_z=\sum_\alpha m_\alpha$.

However, when we have used either of our tracing methods (i) or (ii), 
we have {\it never} found {\it any} such negative parallel windings
up to the highest temperature we
have simulated, $T=1.6J_\perp$.  This has motivated us to define 
a third tracing scheme: (iii) we first search through all possible 
connections to find any paths with $R_{\alpha z} <0$.

\begin{figure}
\epsfxsize=3.2truein
\epsfbox{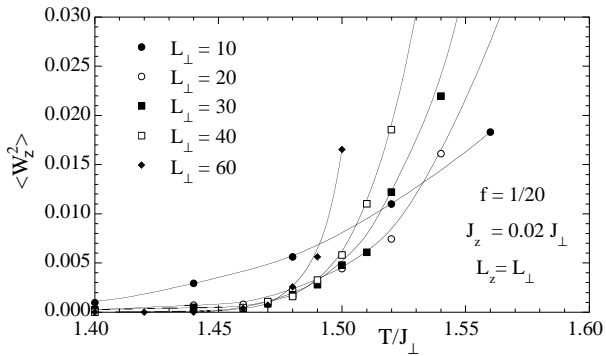}
\caption{Plot of z axis winding $\langle W_z^{2}\rangle$ vs. $T/J_\perp$ for
$L_{\perp}=10$, $20$, $30$, $40$, and $60$, with vortex density $f=1/20$,
anisotropy $J_z=0.02 J_\perp$, and aspect ratio $L_z=L_\perp$.
$\langle W_z^{2}\rangle$ is computed using method (iii), i.e. first find 
all loops that percolate in the negative $\hat z$ direction.  
Solid lines are guides to the eye only.
\label{f3.1}}
\end{figure}

In Fig.\,\ref{f3.1} we show results, using tracing scheme (iii), 
for $\langle W_z^2\rangle$ vs. 
$T/J_\perp$ for the same system parameters and sizes as used  in Fig.\,\ref{f2} 
for $\langle W^2\rangle$.  Note that the values of $\langle W_z^2\rangle$
where the curves for different $L_\perp$ intersect are exceedingly 
small.  The intersection points appear to decrease in $T$ as $L_\perp$
increases, however we do not have the accuracy to make any firm 
conclusions.

\begin{figure}
\epsfxsize=3.2truein
\epsfbox{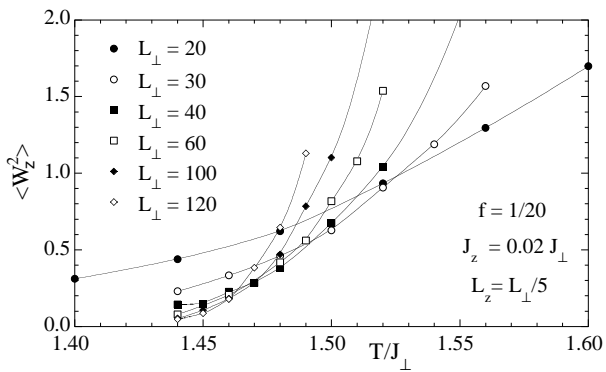}
\caption{Plot of z axis winding $\langle W_z^{2}\rangle$ vs. $T/J_\perp$ for
$L_{\perp}=20$, $30$, $40$, $60$, $100$ and $120$, with vortex density $f=1/20$,
anisotropy $J_z=0.02 J_\perp$, and aspect ratio $L_z=L_\perp/5$.
$\langle W_z^{2}\rangle$ is computed using method (iii), i.e. first find 
all loops that percolate in the negative $\hat z$ direction.  
Solid lines are guides to the eye only.
\label{f3.2}}
\end{figure}

In an attempt to improve the analysis of $\langle W_z^2\rangle$ we have  
repeated the calculation, but using a new system aspect ratio of 
$L_z=L_\perp/5$.  This has the effect of increasing the value where the 
curves of $\langle W_z^2\rangle$ intersect, and so hopefully improving
our accuracy. We have explicitly checked that changing the aspect
ratio does not shift the transition temperature $T_{\Phi}$ 
that is observed in $\langle W^2\rangle$ (see also Section IV.B).
In Fig.\,\ref{f3.2} we show results for $\langle W_z^2\rangle$ vs. 
$T/J_\perp$ for this new aspect ratio.  Again we find no common
intersection point for the sizes considered.  As $L_\perp$ increases, 
the intersection point continues to decrease.  Whether this is a 
failure of the scaling hypothesis of Eq.\,(\ref{eqscalz}), or whether
we have simply failed to reach the scaling limit of sufficiently
large $L_z$ ($L_z=24$ is the largest value in Fig.\,\ref{f3.2}),
we cannot be certain.  Note that in both Figs.\,\ref{f3.1} and 
\ref{f3.2}, $\langle W_z^2\rangle$ appears to be vanishing at a
temperature noticeably {\it above} the $T_\Phi\simeq 1.4J_\perp$ 
where the curves of the
transverse winding, $\langle W^2\rangle$, intersect. 

We have also tried to fit the data of Fig.\,\ref{f3.2} to the scaling 
form, $\langle W_z^2\rangle \sim L^{-x} f_z(tL_\perp^{1/\nu})$, assuming 
a non trivial anomalous scaling dimension $x$ (although we have no
specific theoretical reason to propose this form).  When we do so,
we obtain $T_c=1.44$, $\nu=0.76$, and $x=1.185$, however our data in 
the vicinity of this $T_c$ is too scattered for us to place much
significance on this fit.

Having used tracing method (iii) to first eliminate all possible lines 
percolating in the negative $\hat z$ direction, we can then go and search for 
all possible transversely percolating lines and compute the resulting 
transverse winding $\langle W^2\rangle$.  When we do this, we find our
results for $\langle W^2\rangle$ 
virtually unchanged from tracing method (ii) in the vicinity
of $T_\Phi\simeq 1.4J_\perp$.  The extremely low number 
of negative $\hat z$ percolating lines at this temperature produces no
noticeable effect on the transverse tracing.


\section{Percolating Loops}

\subsection{Summary of Sudb{\o}'s Method}

As discussed in the preceding Section III.A, a transition at $T_{\Phi}$ 
would mark the appearance of infinite transverse loops, as $T$ is increased. 
The idea to explicitly look for transverse paths that percolate 
across the system was first put forth by Jagla and 
Balseiro.\cite{R16}
Later, Sudb{\o} and co-workers\cite{R6,R7,R8} refined this idea.  
They defined a quantity which they 
denoted $O_{L}$, which is the probability that a vortex path exists 
which travels completely across the system in a direction transverse 
to the applied magnetic field, without ever traveling completely across the 
system in the direction parallel to the field.  
If such a path exists in a given configuration, 
that configuration counts as unity in the average for $O_{L}$; if not, 
that configuration counts as zero.  

Since having $W^{2}>0$ in a given configuration necessarily implies that 
there is a percolating transverse loop in that configuration, there is
a close connection between the quantities $\langle W^{2}\rangle$ and 
$O_{L}$.  They differ in that (i) for a configuration with $W^{2}>1$,
and hence with more than one percolating transverse loop, 
the contribution to $O_{L}$ remains unity, rather than increasing with 
the number of percolating transverse loops; and (ii) in a configuration 
with two percolating but oppositely oriented transverse loops, the 
contribution to $O_{L}$ will be unity, but these loops cancel 
each other in their contribution to $W^{2}$, which might therefore be 
zero.

Since $O_{L}$ is a pure number
one might expect it to be a scale invariant quantity, and hence,
similar to $\langle W^{2}\rangle$, plots 
of $O_{L}$ vs. $T$ for different system sizes $L_{\perp}$ should have a 
common intersection point at $T_{\Phi}$.

\begin{figure}
\epsfxsize=3.2truein
\epsfbox{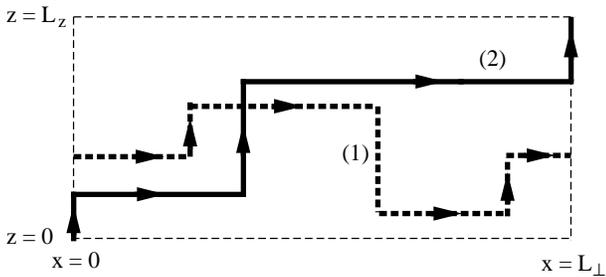}
\caption{Examples of percolating transverse vortex paths. Path (1) 
closes upon itself without ever winding about the system in the $\hat 
z$ direction.  Path (2) only closes upon itself after winding once in 
the $\hat z$ direction.  Both paths contribute to the calculation of
$O_{L}^{\prime}$ by method (ii$^{\prime}$).  Only path (1) contributes
to the calculation of $O_{L}$ by method (ii).  The thin dashed lines 
represent the periodic boundaries of the system.
\label{f4}}
\end{figure}

Sudb{\o} and co-workers' method of searching for such 
percolating transverse paths is similar 
to our method (ii) except for one crucial difference.\cite{R17}  They do not 
require that the transverse path closes upon itself; they only 
require that the path start  at one end of the system, say at $x=0$, 
and continue until reaching the opposite end, $x=L_{\perp}$, while 
keeping the distance traveled along $\hat z$ less than $L_{z}$, that is
the displacement traveled along the path $\alpha$ satisfies
$R_{x\alpha}=L_{\perp}$ and $R_{z\alpha}<L_{z}$.  
Since, by the periodic boundary conditions, all paths {\it must} 
eventually close upon themselves, there are two possibilities for 
the transverse percolating paths that Sudb{\o} and co-workers find.
We illustrate these in Fig.\,\ref{f4}: 
(1) the path $\alpha$ eventually closes upon itself without ever 
traveling the length $L_{z}$, in which case $R_{z\alpha}=0$; or (2) the 
path $\alpha$, when followed until it closes upon itself, does 
wind up traveling the length $L_{z}$, with a displacement 
$R_{z\alpha}=L_{z}$; in this case, our method (ii) would consider this 
path as part of the field induced vortex lines $n_{s\mu}^\mathrm{lines}$, 
contributing to the winding ${\bf W}$, rather than as a transverse loop 
that contributes to $n_{s\mu}^\mathrm{loops}$.  We will call Sudb{\o}'s 
path tracing method (ii$^{\prime}$) to distinguish it from our method (ii).
The probability for a percolating path using method (ii$^{\prime}$)
we will denote by $O_{L}^{\prime}$; using method (ii) we will denote 
it by $O_{L}$.  Paths of type (2) will contribute to 
$O_{L}^{\prime}$, but not to $O_{L}$.  
We will see that there are very dramatic differences 
between these two methods, and that only $O_{L}$ gives 
self-consistent results.

\subsection{Numerical Results}

First, we note that if we use method (i) (random connections)
to search for percolating 
transverse paths, the result is essentially the same as found for 
$\langle W^2\rangle$ in section III.B.  As $L_{\perp}$ increases,
the probability  to find a percolating transverse loop steadily
decreases for the entire temperature range,
becoming immeasurably small for our biggest system size.  Hence we 
will focus here on methods (ii) and (ii$^{\prime}$).

\begin{figure}
\epsfxsize=3.2truein
\epsfbox{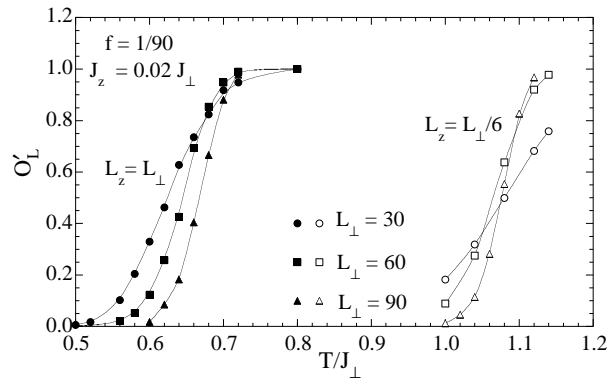}
\caption{Plot of percolation probability $O_{L}^{\prime}$ vs. 
$T/J_{\perp}$ for
$L_{\perp}=30$, $60$, and $90$, with vortex density $f=1/90$, and
anisotropy $J_z=0.02 J_\perp$.   Solid symbols on the left are for
aspect ratio $L_z=L_\perp$; open symbols on the right are for
aspect ratio $L_z=L_\perp/6$.
$O_{L}^{\prime}$ is computed using method (ii$^{\prime}$), following
Sudb{\o} and co-workers.  Solid lines are guides to the eye only.
\label{f5}}
\end{figure}

We now consider the computation of $O_{L}^{\prime}$ using 
method (ii$^{\prime}$), the one used by Sudb{\o} 
and co-workers, which never checks to see how the percolating 
transverse path closes upon itself.  We first use  parameters 
$J_{z}=0.02 J_{\perp}$ and $L_{z}=L_{\perp}$, the same as in section 
III.B, but a more dilute density of vortex lines $f=1/90$.  
For these parameters, the vortex lattice melting temperature
is $T_{m}\simeq 0.49 J_{\perp}$, and the zero field critical
temperature, as before, is $T_{c0}\simeq 1.14 J_{\perp}$. 
These parameters are 
very close to the parameters of one of the cases studied by Sudb{\o} 
and co-workers in Refs.~\onlinecite{R7} and \onlinecite{R8} (they used $f=1/90$, 
$J_{z}=(1/49) J_{\perp}$ and $L_{z}\sim L_{\perp}$).  Our results
for $O_{L}^{\prime}$ vs $T/J_\perp$, for three system sizes, $L_{\perp}=30$, 
$60$ and $90$, are shown as the solid symbols on the left hand side of 
Fig.\,\ref{f5}.  These results agree quite closely with those 
of Sudb{\o} and co-workers (see Fig.\,8 of Ref. \onlinecite{R7}),
and seem to show what might be a common 
intersection of the three curves near $T\simeq 0.7 J_{\perp}$.  However, we 
now consider the same parameters and sizes $L_{\perp}$, 
only using a different system aspect ratio, $L_{z}=L_{\perp}/6$.  The results 
are shown as the open symbols on the right hand side of Fig.\,\ref{f5}.
We see that there no longer appears to be a common intersection 
point, but more importantly, the curves have all shifted dramatically 
to higher temperatures.  Thus, any value of $T_{\Phi}$ that one might 
try to extract from $O_{L}^{\prime}$ depends sensitively on the system 
aspect ratio.  We have also considered other values of the aspect
ratio $L_{z}/L_{\perp}$, not shown here.  The clear trend is that the 
sharp rise in $O_{L}^{\prime}$ shifts to increasing temperatures as
$L_{z}/L_{\perp}$ decreases.
But if $T_{\Phi}$ represents a true phase transition,
it must be {\it independent} of aspect ratio.  We therefore conclude that
$O_{L}^{\prime}$ and method (ii$^{\prime}$) do not give any 
self-consistent evidence of the proposed vortex loop blowout transition.

\begin{figure}
\epsfxsize=3.2truein
\epsfbox{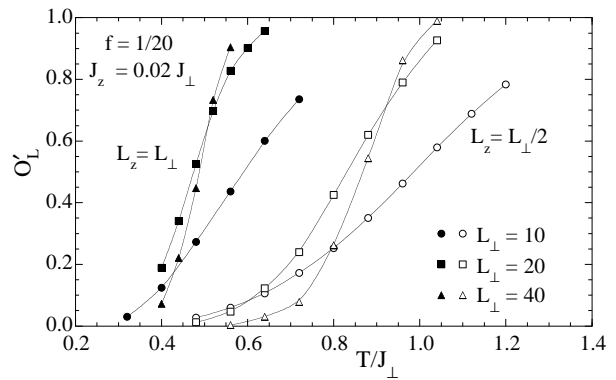}
\caption{Plot of percolation probability $O_{L}^{\prime}$ vs. 
$T/J_{\perp}$ for
$L_{\perp}=10$, $20$, and $40$, with vortex density $f=1/20$, and
anisotropy $J_z=0.02 J_\perp$.   Solid symbols on the left are for
aspect ratio $L_z=L_\perp$; open symbols on the right are for
aspect ratio $L_z=L_\perp/2$.
$O_{L}^{\prime}$ is computed using method (ii$^{\prime}$), following
Sudb{\o} and co-workers.  Solid lines are guides to the eye only.
\label{f6}}
\end{figure}

The problems with $O_{L}^{\prime}$ are even clearer if we consider 
the parameters $f=1/20$ and $J_{z}=0.02 J_{\perp}$, the same ones 
used for our computation of $\langle W^{2}\rangle$ in section III.B.
In Fig.\,\ref{f6} we show our results for $L_{\perp}=10$, $20$, and 
$40$, for the two aspect ratios $L_{z}=L_{\perp}$ and $L_{z}=L_{\perp}/2$.
In both cases, there is no common intersection point of the three 
curves for the three sizes, and the curves for the smaller aspect ratio 
are shifted to higher temperatures.  Note also that, for both aspect
ratios, the temperatures 
at which $O_{L}^{\prime}$ rises to unity lie quite significantly below 
the value of $T_{\Phi}\simeq 1.4 J_{\perp}$ found from our analysis of
$\langle W^{2}\rangle$.  

\begin{figure}
\epsfxsize=3.2truein
\epsfbox{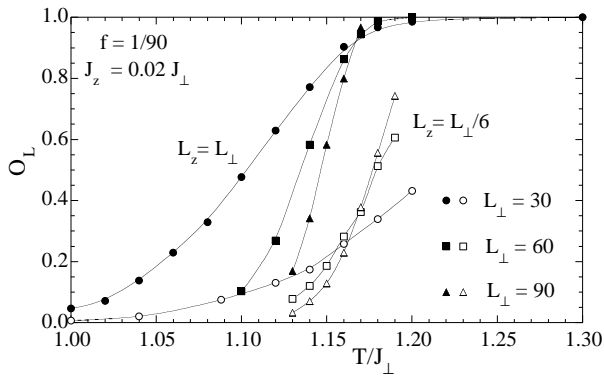}
\caption{Plot of percolation probability $O_{L}$ vs. $T/J_{\perp}$ for
$L_{\perp}=30$, $60$, and $90$, with vortex density $f=1/90$, and
anisotropy $J_z=0.02 J_\perp$.   Solid symbols are for
aspect ratio $L_z=L_\perp$; open symbols are for
aspect ratio $L_z=L_\perp/6$.
$O_{L}$ is computed using method (ii), where all loops close upon 
themselves.  All curves approach a common intersection point, 
$T_{\Phi}\simeq 1.17 J_{\perp}$, independent of the aspect ratio.  
Solid lines are guides to the eye only.
\label{f7}}
\end{figure}

We next consider the the computation of $O_{L}$ using 
method (ii) (percolating transverse path must close upon itself keeping 
$R_{z\alpha}=0$).  In Fig.\,\ref{f7} we show results using the same
parameters as were used to compute $O_{L}^{\prime}$in Fig.\,\ref{f5},
i.e. $f=1/90$, $J_{z}=0.02 J_{\perp}$, and $L_{\perp}=30$, $60$ and $90$,
for the same two aspect ratios $L_{z}=L_{\perp}$ and $L_{z}=L_{\perp}/6$.
We see now that for both aspect ratios, curves for the three 
different sizes appear to approach a common intersection point, 
$T_{\Phi}\simeq 1.17 J_{\perp}>T_{c0}\simeq 1.14 J_{\perp}$, and 
that this intersection point is independent of aspect ratio (note: for 
$L_{z}=L_{\perp}/6$, the thinness of the system $L_z=5$, for 
$L_{\perp}=30$, presumably makes it too small to be in the scaling 
region, hence it intersects the other two curves at somewhat lower 
temperatures).

\begin{figure}
\epsfxsize=3.2truein
\epsfbox{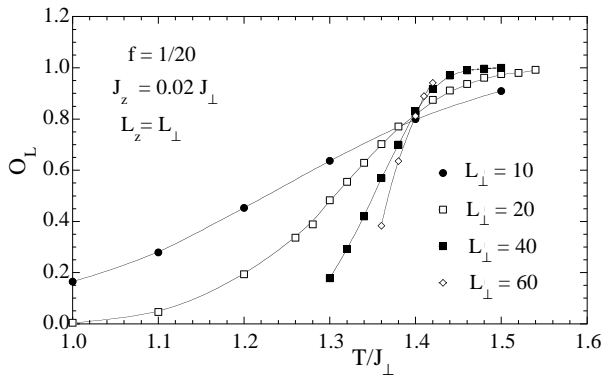}
\caption{Plot of percolation probability $O_{L}$ vs. $T/J_{\perp}$ for
$L_{\perp}=10$, $20$, $40$, and $60$, with vortex density $f=1/20$, 
anisotropy $J_z=0.02 J_\perp$, and aspect ratio $L_z=L_\perp$.
$O_{L}$ is computed using method (ii), where all loops close upon 
themselves.  All curves approach a common intersection point, 
$T_{\Phi}\simeq 1.4 J_{\perp}$, in agreement with the analysis of $\langle 
W^{2}\rangle$ in Fig.\,\ref{f2}.
Solid lines are guides to the eye only.
\label{f8}}
\end{figure}

In Fig.\,\ref{f8} we show similar results using the same parameters
as were used in our computation of $\langle W^{2}\rangle$ in section 
III.B, i.e. $f=1/20$, $J_{z}=0.02 J_{\perp}$ and $L_{z}=L_{\perp}$.
We see that the curves of $O_{L}$ vs. $T/J_\perp$ for the different system sizes, 
$L_{\perp}=10$, $20$, $30$ and $40$, all intersect at a common 
point, $T_{\Phi}\simeq 1.4 J_{\perp}$.  This is exactly the same value as found 
in our analysis of $\langle W^{2}\rangle$ (see Fig.\,\ref{f2}).

We therefore conclude that $O_{L}$ gives a self-consistent
determination of $T_{\Phi}$, and that this value is considerably 
larger than estimates one would get from consideration of $O_{L}^{\prime}$.
In fact, estimates of a $T_{\Phi}$ from $O_{L}^{\prime}$ all lie {\it below}
the zero field critical temperature $T_{c0}$ and {\it decrease} as $f$ 
{\it increases}, while the  
values determined from $O_{L}$ all lie {\it above} $T_{c0}$, and {\it 
increase} as $f$ {\it increases}.

\begin{figure}
\epsfxsize=3.2truein
\epsfbox{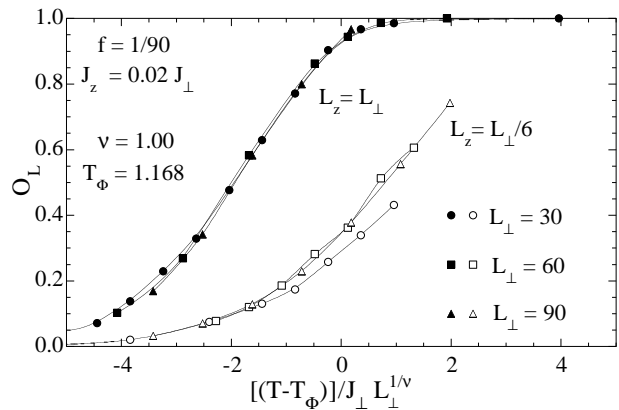}
\caption{Scaling collapse $O_{L}$ vs. $[(T-T_{\Phi})/J_{\perp}]L_{\perp}^{1/\nu}$
of data from Fig.\,\ref{f7}, for $f=1/90$.
A reasonably good collapse is found 
for all sizes $L_{\perp}$, for both aspect ratios $L_{z}/L_{\perp}$, 
using a single value of $T_{\Phi}\simeq 1.168 J_{\perp}$ and $\nu = 1$.
Solid lines are guides to the eye only.
\label{f9}}
\end{figure}

If $O_{L}$ is indeed a scale invariant quantity, we can 
postulate that it should obey a scaling relation similar to $\langle 
W^{2}\rangle$, i.e.,
\begin{equation}
    O_{L}(T,L_{\perp})=\tilde{f}(tL^{1/\nu})
\label{eqScaleO}
\end{equation}
Based on our analysis of $\langle W^{2}\rangle$ in section III.B, we 
may expect $\nu\simeq 1$.  In Fig.\,\ref{f9} we therefore show a 
scaling collapse of the data for $f=1/90$ from Fig.\,\ref{f7},
plotting $O_{L}$ vs. 
$[(T-T_{\Phi})/J_{\perp}]L_{\perp}$, where $T_{\Phi}$ is determined by a best 
fit of the data to the scaling form.  We find a reasonably good 
collapse
for all sizes, for both aspect ratios, using a single value of 
$T_{\Phi}\simeq 1.168 J_{\perp}$.

\begin{figure}
\epsfxsize=3.2truein
\epsfbox{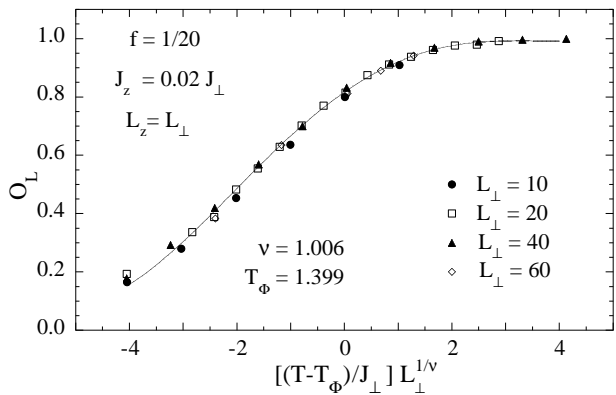}
\caption{Scaling collapse $O_{L}$ vs. $[(T-T_{\Phi})/J_{\perp}]L_{\perp}^{1/\nu}$
of data from Fig.\,\ref{f8}, for $f=1/20$.
An excellent collapse is found 
for all sizes $L_{\perp}$, using a value of $T_{\Phi}\simeq 1.399 
J_{\perp}$ 
and $\nu = 1.006$.  These values agree well with those obtained from 
the scaling collapse of $\langle W^{2}\rangle$.  The solid line is the
fitted polynomial curve.
\label{f10}}
\end{figure}

In Fig.\,\ref{f10} we show a similar scaling collapse of the data
for $f=1/20$ from Fig.\,\ref{f8}.  Fitting the data for $O_{L}$ to a
4th order polynomial expansion of the scaling function, we find an 
excellent collapse, for all system sizes, using the parameters 
$T_{\Phi}\simeq 1.399 J_{\perp}$ and $\nu = 1.006$.  These results agree
very well with the values obtained from the scaling analysis of
$\langle W^{2}\rangle$, given in section III.B.  The quality of the
collapse is very much better here than it was for $\langle W^{2}\rangle$
in Fig.\,\ref{f3}.

\begin{figure}
\epsfxsize=3.2truein
\epsfbox{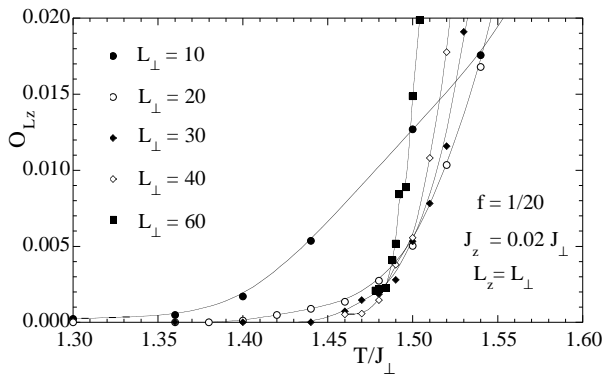}
\caption{Plot of percolation probability $O_{Lz}$ vs. $T/J_{\perp}$ for
$L_{\perp}=10$, $20$, $30$, $40$, and $60$, with vortex density $f=1/20$, 
anisotropy $J_z=0.02 J_\perp$, and aspect ratio $L_z=L_\perp$.
$O_{Lz}$ is computed using tracing method (iii).  
Solid lines are guides to the eye only.
\label{f10.1}}
\end{figure}

\begin{figure}
\epsfxsize=3.2truein
\epsfbox{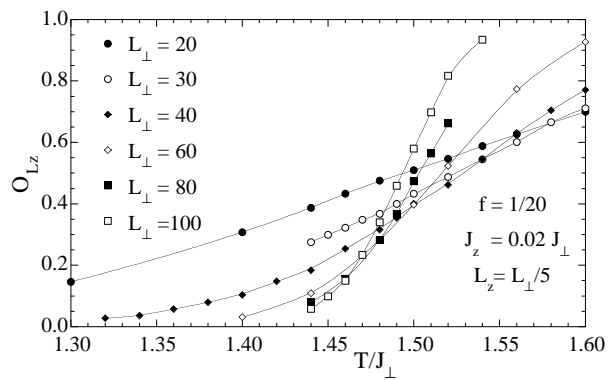}
\caption{Plot of percolation probability $O_{Lz}$ vs. $T/J_{\perp}$ for
$L_{\perp}=20$, $30$, $40$, $60$, $80$, and $100$, 
with vortex density $f=1/20$, 
anisotropy $J_z=0.02 J_\perp$, and aspect ratio $L_z=L_\perp/5$.
$O_{Lz}$ is computed using tracing method (iii).  
Solid lines are guides to the eye only.
\label{f10.2}}
\end{figure}

Finally, in analogy with the winding $\langle 
W_z^2\rangle$, we have also considered the probability $O_{Lz}$
to find a vortex path percolating through the system in the negative  $\hat z$ 
direction, opposite to the applied magnetic field.  We expect
$O_{Lz}$ to obey a scaling relation similar to that of 
Eq.\,(\ref{eqScaleO}).  To  compute
$O_{Lz}$ we have used tracing method (iii) in which we explicitly
search through all possible connections to find any such paths. 
We show our results for $O_{Lz}$ vs. $T/J_{\perp}$ for vortex 
density $f=1/20$ and anisotropy $J_{z}=0.02 J_{\perp}$ in 
Figs.\,\ref{f10.1} and \ref{f10.2}, for system aspect ratios
$L_{z}=L_{\perp}$ and $L_{z}=L_{\perp}/5$ respectively.  As with
$\langle W_z^{2}\rangle$ shown in Figs.\,\ref{f3.1} and \ref{f3.2},
the intersection points of the curves for different sizes
appear to decrease in temperature as $L_{\perp}$ increases.
Again, we cannot say if this is a failure of our scaling hypothesis,
or a failure to reach sufficiently large
$L_{z}$.  Also, analogous to our findings for the windings $\langle 
W^{2}\rangle$ and $\langle W_{z}^{2}\rangle$,
$O_{Lz}$ appears to be vanishing at a temperature {\it above} the
$T_{\Phi}=1.4J_{\perp}$ where the curves of $O_{L}$ intersect.

%


\section{Specific Heat}

If $T_{\Phi}$, as determined by $\langle W^{2}\rangle$ or $O_{L}$, 
does indeed represent a true thermodynamic transition, we would 
expect to see some signature of this transition in more conventional 
thermodynamic quantities.  In the recent experiments of 
Ref.\onlinecite{R9}, 
a step like anomaly in the specific heat $C$ was observed in the vortex 
line liquid region, reminiscent of an inverted mean field transition.
In their numerical simulations,\cite{R7,R8} Nguyen and Sudb{\o} claimed to see 
an anomalous glitch in the specific heat at the temperature they 
identified as $T_{\Phi}$ from their calculation of $O_{L}^{\prime}$.
However, this glitch corresponded to only a single data point very 
slightly displaced above an otherwise smooth background; and in the 
previous section we have demonstrated that $O_{L}^{\prime}$ 
significantly underestimates $T_{\Phi}$, hence there is no reason to 
expect any anomaly in $C$ at that temperature.

In this section we report on high precision measurements of the 
specific heat $C$, for the same parameters we have studied in the 
earlier sections.  If the $T_{\Phi}$, as found using the vortex path
tracing method (ii), is 
indeed a true thermodynamic phase transition with critical 
exponent $\nu\simeq 1$ 
(as our scaling analyses found), then hyperscaling would suggest a 
specific heat exponent of $\alpha = 2 - d\nu \simeq -1$.  We thus
do not expect to see a diverging $C$, however some feature should
be present.

\begin{figure}
\epsfxsize=3.2truein
\epsfbox{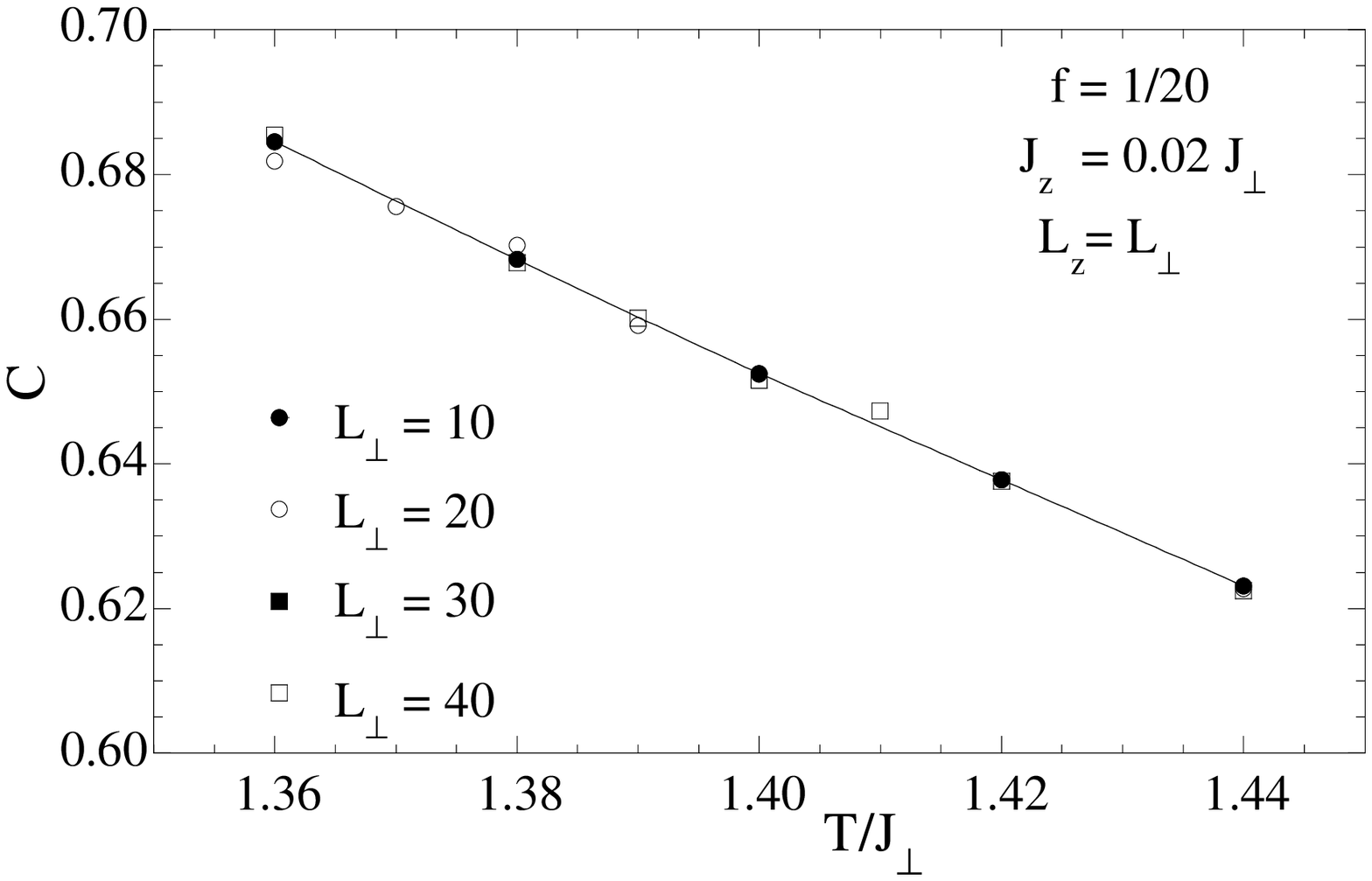}
\caption{Specific heat $C$ vs. $T/J_{\perp}$ for vortex density $f=1/20$,
anisotropy $J_{z}=0.02 J_{\perp}$, and aspect ratio 
$L_{z}=L_{\perp}$, for system sizes $L_{\perp} =10$, $20$, $30$ and 
$40$.  No hint of any anomaly is found near the previously determined 
$T_{\Phi}\simeq 1.4 J_{\perp}$.  The solid line is a guide to the eye only.
\label{f11}}
\end{figure}

In Fig.\,\ref{f11} we plot $C$ vs. $T/J_\perp$, in the near vicinity
of $T_{\Phi}\simeq 1.4J_{\perp}$, for the same
parameters $f=1/20$, $J_{z}=0.02 J_{\perp}$, and $L_{z}=L_{\perp}$ as  
used in Figs.\,\ref{f2}, \ref{f8} and \ref{f10}.  
We show results for $L_{\perp}=10$, $20$, $30$ and $40$, using from 
$3 - 10 \times 10^{7}$ Monte Carlo passes through the lattice,
depending on the system size.  
We find no noticeable finite size dependence, and no
hint of any feature at all, near the previously determined $T_{\Phi}\simeq 
1.4 J_{\perp}$.

\begin{figure}
\epsfxsize=3.2truein
\epsfbox{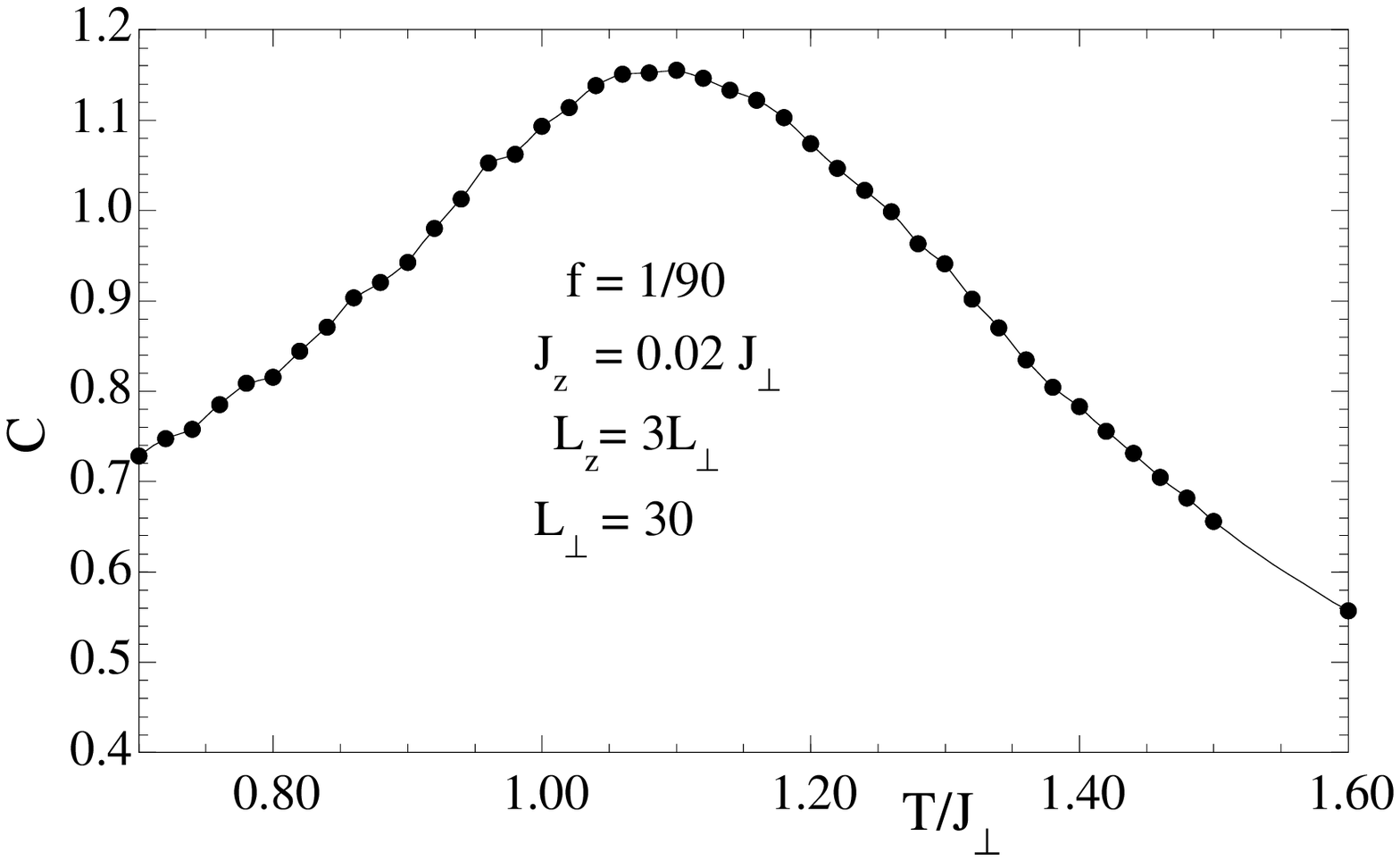}
\caption{Specific heat $C$ vs. $T/J_{\perp}$ for vortex density $f=1/90$,
anisotropy $J_{z}=0.02 J_{\perp}$, and aspect ratio 
$L_{z}=3L_{\perp}$, for system size $L_{\perp} =30$.  
No hint of any anomaly is found near the previously determined 
$T_{\Phi}\simeq 1.168 J_{\perp}$.  The solid line is a guide to the eye only.
\label{f12}}
\end{figure}

In Fig.\,\ref{f12} we plot $C$ vs. $T/J_\perp$, over a broad temperature 
range, for the same parameters
$f=1/90$ and $J_{z}=0.02 J_{\perp}$ as used in Figs.\,\ref{f7} and
\ref{f9}, but for a single large system size $L_{\perp}=30$ and 
$L_{z}=90$.  Again we see no hint of any anomaly near the previously 
determined $T_{\Phi}\simeq 1.168 J_{\perp}$.

\section{Conclusions}

We have carried out detailed Monte Carlo investigations of the
3D uniformly frustrated XY model in order to 
search for a proposed ``vortex loop blowout'' transition within the 
vortex line liquid phase of a pure extreme type-II superconductor.
Such a transition had been predicted from general theoretical arguments 
by Te\v{s}anovi\'{c}.\cite{R5}  Evidence for such a transition was claimed in 
numerical simulations by Sudb{\o} and co-workers,\cite{R6,R7,R8} 
and in specific heat 
measurements on high purity YBCO single crystals.\cite{R9}  
We have made explicit measurements of the vortex line windings
$\langle W^{2}\rangle$ and $\langle W_{z}^{2}\rangle$, which are the
key quantities in Te\v{s}anovi\'{c}'s theory.  We have re-examined
Sudb{\o}'s calculation of the percolation probability $O_L$.

Our results raise several questions concerning Te\v{s}anovi\'{c}'s 
theory.  We have found that the values of $\langle W^{2}\rangle$ and 
$\langle W_{z}^{2}\rangle$ depend sensitively on the precise scheme one uses 
to trace out vortex line paths.  For the natural choice of random 
connectivity at vortex line intersections, both $\langle W^{2}\rangle$ and 
$\langle W_{z}^{2}\rangle$ appear to vanish at all temperatures as 
$L\to\infty$.  Only when we specifically search first for percolating paths, 
when computing the windings, do we find that the windings converge to 
non zero values above a certain temperature.  In this case, we find 
that the transverse winding $\langle W^{2}\rangle$ obeys the 
finite size scaling form expected from Te\v{s}anovi\'{c}'s theory,
however the critical exponent we find is $\nu\simeq 1$, rather than
the predicted $\nu_{XY}\sim 2/3$ of the inverted 3D XY transition.  For the 
longitudinal winding $\langle W_{z}^{2}\rangle$ we have been unable 
to find the expected scaling form.  Whether this is because 
$\langle W_{z}^{2}\rangle$ does not scale, or because our systems are all too
small to be in the scaling limit, we cannot be certain.  It does 
appear that, upon cooling, $\langle W_{z}^{2}\rangle$ vanishes at a 
temperature above that where $\langle W^{2}\rangle$ vanishes.  This 
would be contrary to Te\v{s}anovi\'{c}'s theory.  However, since we 
have not succeeded to find scaling for $\langle W_{z}^{2}\rangle$, we 
cannot be certain of knowing exactly where it vanishes as $L\to\infty$.

Independent of Te\v{s}anovi\'{c}'s theory, it is 
natural to think that, as temperature and hence vorticity increases, 
the vortex lines may form percolating paths (note however that
the directedness of the vortex line segments, and the condition of
divergenceless paths, means that this is no ordinary percolation 
problem!).  We have therefore,
following Sudb{\o}  and co-workers,
searched explicitly for such percolating paths in the direction 
transverse to the applied magnetic field, as well as in the direction 
parallel but opposite to the applied magnetic field.  Defining 
transverse percolation as the existence of a vortex line path that 
extends entirely across the system in the direction transverse to the 
applied magnetic field {\it without} simultaneously extending entirely 
across the system in the parallel direction, we have shown that 
Sudb{\o}'s procedure, which ignores the transverse periodic boundary 
conditions and does not require the percolating path to close upon 
itself, leads to inconsistent predictions for the transition 
temperature as one varies the system aspect ratio.  Only by requiring 
that the transverse percolating path close upon itself, {\it without} 
ever winding in the parallel direction, do we find a consistent 
transition temperature independent of aspect ratio.  The 
percolation transition found this way agrees both in critical 
temperature $T_{\Phi}$ and exponent $\nu$ with the results from
our analysis of the transverse winding $\langle W^{2}\rangle$.
We have also computed the probability to find a percolating path in 
the direction parallel but opposite to the applied magnetic field.
Here, analogous to our results for $\langle W_{z}^{2}\rangle$, this
negative $\hat z$ percolation appears to occur at a temperature higher than 
that of the transverse percolation, however we have not succeeded to 
find a clear scaling of this parallel percolation probability.

Note that the transverse percolation transition temperature
$T_{\Phi}(f)$ that we find {\it increases} above the zero field
transition temperature $T_{c0}$ as the magnetic flux density $f$ {\it 
increases}.  This is in striking contrast to the conclusion of 
Sudb{\o} and co-workers who proposed $T_{\Phi}(f)$ to {\it decrease} 
below $T_{c0}$ as $f$ {\it increases}.  

While our results do seem consistent with a well defined transverse 
percolation transition, one can ask if this is a purely geometrical 
feature of the vortex line paths, or whether it also corresponds to a 
true thermodynamic phase transition, i.e. something one could detect 
in a suitable thermodynamic derivative of the free energy.  To 
investigate this question we have carried out high precision Monte 
Carlo measurements of the specific heat $C$. Our results 
for $C$ show no feature whatsoever near the percolation 
transition $T_{\Phi}$, nor do we find any finite size effect.  In 
particular we see no evidence for a step like feature as was 
observed experimentally in YBCO.

To conclude, we have found evidence for a well defined transverse 
percolation temperature within the vortex line liquid phase of a 
model type-II superconductor.  The 
connection between this transition and Te\v{s}anovi\'{c}'s theory of a 
vortex loop ``blowout'' transition remains unclear.
It also remains unclear whether or not this percolation transition has 
any observable thermodynamic manifestation.

\section*{Acknowledgements}

We would like to thank Prof. Z. Te\v{s}anovi\'{c} and Prof. A. Sudb{\o} 
for many helpful conversations.
This work was supported by the
Engineering Research Program of the Office of Basic Energy Sciences
at the Department of Energy grant DE-FG02-89ER14017, the
Swedish Natural Science Research Council Contract No. E 5106-1643/1999,
and by the resources of the Swedish High Performance Computing Center 
North (HPC2N).  Travel between Rochester and Ume{\aa} was supported by 
grants NSF INT-9901379 and STINT 99/976(00).


\begin{thebibliography}{99}

\bibitem{R1} E.~Zeldov {\it et al.}, Nature {\bf 375}, 373 (1995);
A.~Schilling {\it et al.}, Nature {\bf 382}, 791 (1996).
    
\bibitem{R2} M.~V.~Feigel'man, V.~B.~Geshkenbein, L.~B.~Ioffe and 
    A.~I.~Larkin, Phys. Rev. B {\bf 48}, 16641 (1993).
  
\bibitem{R3} Y.-H.~Li and S.~Teitel, Phys. Rev. B {\bf 47}, 359 (1993).
\bibitem{R3.1} Y.-H.~Li and S.~Teitel, Phys. Rev. B {\bf 49}, 4136 (1994).
\bibitem{R3.2} T.~Chen and S.~Teitel, Phys. Rev. B {\bf 55}, 11766 (1997).

\bibitem{R4} X.~Hu, S.~Miyashita and M.~Tachiki, Phys. Rev. Lett.
  {\bf 79}, 3498 (1997) and Phys. Rev. B {\bf 58} 3438 (1998); 
   A.~K.~Nguyen and A.~Sudb{\o}, Phys. Rev. B {\bf 58}, 2802 (1998); 
   P.~Olsson and S.~Teitel, Phys. Rev. Lett. {\bf 82}, 2183 (1999).

\bibitem{R5} Z.~Te\v{s}anovi\'{c}, Phys. Rev. B {\bf 59}, 6449 (1999)
  and Phys. Rev. B {\bf 51}, 16204 (1995).

\bibitem{R6} S.~K.~Chin, A.~K.~Nguyen and A.~Sudb{\o}, Phys. Rev. B 
{\bf 59}, 14017 (1999).
\bibitem{R7} A.~K.~Nguyen and A.~Sudb{\o}, Europhys. Lett. {\bf 46}, 
780 (1999).
\bibitem{R8} A.~K.~Nguyen and A.~Sudb{\o}, Phys. Rev B {\bf 60}, 
15307 (1999).

\bibitem{R9} F.~Bouquet {\it et al.}, Nature {\bf 411}, 448 (2001).

\bibitem{R10} Y.-H.~Li and S.~Teitel, Phys. Rev. Lett. {\bf 66}, 3301 
   (1991).
   
\bibitem{R11} P.~Olsson, Europhys. Lett. {\bf 58}, 705 (2002).

\bibitem{R11.1} For large systems, it is crucial to have an efficient 
algorithm to search for such paths.  We use the following.  First, 
all intersection points are located.  Picking one such point at 
random, we trace out a path starting from this intersection point
until it arrives at another intersection point.  If the height 
traveled from the starting to the new intersection point is $\Delta 
z>0$, we stop this search and start again at a different intersection point;
if not, we continue tracing the  path until then next intersection 
point is encountered and then repeat the height test.  Since most 
paths connect to field lines which travel in the $+z$ direction, most
such tracings are quickly aborted.  However, since each possible 
transverse loop with $R_{z\alpha}=0$ contains an intersection point 
that is at the largest
height of all intersection points on that loop, we are guarenteed 
to ultimately find this path with this search algorithm.
   
\bibitem{R12} C.~Dasgupta and B.~I.~Halperin, Phys. Rev. Lett. {\bf 
47}, 1556 (1981).

\bibitem{R12.1} E.~Fradkin, B.~A.~Huberman, and S.~H.~Shenker, Phys. Rev. B 
{\bf 18}, 4789 (1978); G.~Carneiro, Phys. Rev. B {\bf 45}, 2391 (1992).

\bibitem{R12.2}T.~Chen and S.~Teitel, Phys. Rev. B {\bf 55}, 15197 (1997).

\bibitem{R12.3} M.~Kiometzis, H.~Kleinert, and A.~M.~J.~Schakel, 
Fortschr. Phys. {\bf 43}, 697 (1995).

\bibitem{R13} P.~Olsson and S.~Teitel, Phys. Rev. Lett. {\bf 80}, 1964 
(1998).

\bibitem{R14} D.~R.~Nelson, Phys. Rev. Lett. {\bf 60}, 1973 (1988); 
J. Stat. Phys. {\bf 57}, 511 (1989); D.~R.~Nelson and H.~S.~Seung, Phys. 
Rev. B {\bf 39}, 9153 (1989). 


\bibitem{R16} E.~A.~Jagla and C.~A.~Balseiro, Phys. Rev. B {\bf 53}, 
R538 (1996); {\it ibid.} {\bf 53}, 15305.  In these works, the authors 
considered {\it all} transverse percolating loops, including those with net 
winding in the parallel direction, $R_{\alpha z}>0$.  One can show 
that the onset of such loops, which in general involve the 
participation of the field induced lines and so do have $R_{\alpha z}>0$, 
coincides with the vanishing of the 
longitudinal helicity modulus and so occurs at the melting $T_{m}$,
rather than the proposed $T_{\Phi}$.  Only by restricting to loops
with $R_{\alpha z}=0$ does one probe $T_{\Phi}$.

\bibitem{R17} A.~Sudb{\o}, private communication.
\end{thebibliography}
\end{document}